\documentclass[11pt]{article}  

\usepackage{graphicx,color,epsf} 

\begin{document}

\begin{center}
{\LARGE \bf Thermodynamics of the scalar radiation \\
in the presence of a reflecting plane wall}
\\ 
\vspace{1cm} 
{\large E. S. Moreira Jr.}
\footnote{E-mail: moreira@unifei.edu.br}   
\\ 
\vspace{0.3cm} 
{\em Instituto de Matem\'{a}tica e Computa\c{c}\~{a}o,}  
{\em Universidade Federal de Itajub\'{a},}   \\
{\em Itajub\'a, Minas Gerais 37500-903, Brazil}

\vspace{0.3cm}
{\large October, 2018}
\end{center}
\vspace{1cm}


\abstract{ 
This paper investigates further how the presence of a single reflecting plane wall
modifies the usual Planckian forms in the thermodynamics of the massless scalar radiation in $N$-dimensional Minkowski spacetime. This is done in a rather
unconventional way by integrating the energy density over space to obtain
the internal energy and from that
the Helmholtz free energy. The reflecting wall is modelled by assuming
the Dirichlet or the Neumann boundary conditions on the wall.  
It is found that when $N>2$ integration over 
space eliminates dependence on the curvature coupling parameter $\xi$. 
Unexpectedly though,
when $N=2$, the internal energy and the corresponding thermodynamics turn out to be dependent on $\xi$. For instance, the correction to the two-dimensional Planckian heat capacity is $\mp \xi
k_{B}$ (minus for Dirichlet, plus for Neumann). Other aspects of this dependence on $\xi$ are also discussed. Results are confronted with those 
in the literature concerning related setups of reflecting walls (such as slabs) where conventional (i.e., global) approaches to obtain thermodynamics have been used.
}

\vspace{0.5cm}
\hspace{-0.5cm} \emph{Keywords:} hot scalar radiation, boundaries

\section{Introduction}
\label{int}
Scalar fields are common ingredients of models in particle physics,
cosmology and gravitation. It is therefore pertinent to investigate the many aspects of thermodynamics of the scalar radiation. The traditional way of obtaining the equations of state of blackbody radiation is by means of statistical mechanics of standing waves in a cavity. 
There is however another approach which is to use the ensemble average  $\left<T^\mu{}^\nu\right>$
of the stress-energy-momentum tensor.
Applying tools of quantum field theory at finite temperature $T$, 
one finds for a massless scalar field in four-dimensional Minkowski spacetime the following homogeneous and isotropic $\left<T^\mu{}^\nu\right>$,
\begin{equation}
\left<T^\mu{}^\nu\right> ={\rm diag}(\rho,p,p,p),
\label{semt}
\end{equation}
where
\begin{eqnarray}
&&
\rho=\frac{\pi^2}{30}\frac{(k_{B}T)^4}{(\hbar c)^3},
\hspace{1cm}
p=\frac{\rho}{3}.
\label{gas}
\end{eqnarray}
The internal energy $U$ of the scalar radiation in a cavity of volume $V$
can be obtained by integrating over space the energy density $\rho$ in 
eq. (\ref{gas}), 
\begin{equation}
U=\frac{\pi^2}{30}V\frac{(k_{B}T)^4}{(\hbar c)^3}.
\label{ie}
\end{equation}
The expression in eq. (\ref{ie}) and the pressure $p$ in eq. (\ref{gas}) are the familiar equations of state for the scalar blackbody radiation
in a cavity of volume $V$ and  at thermodynamic equilibrium with the cavity walls at temperature $T$. (For electromagnetic radiation  $U$ is twice that in eq. (\ref{ie}), see e.g. ref. \cite{hua87}.)

Nevertheless, eq. (\ref{ie}) is an approximation which holds only
when the temperature is high enough or the cavity is big enough.
This follows from the fact that $\left<T^\mu{}^\nu\right>$ is not
homogeneous and isotropic when the walls of the cavity are taking into account. Indeed, as has been shown in ref. \cite{ken80}, in the presence of a single reflecting plane wall $\rho$ is  given as in eq. (\ref{gas}) far away from the wall;
but it is radically modified when the wall is approached.

The main purpose of this paper is to calculate the internal energy $U$ as described above,
i.e., integrating $\rho$ over space,
now considering the inhomogeneous energy density $\rho$
in the presence of a single reflecting plane wall \cite{ken80,tad86,mor15} (in so doing, effects due to other walls of the cavity will assumed to be negligible).
Before embarking on this task, which will be referred in this work as the
local procedure to determine $U$,
a short account on early works 
is required.

In order to obtain the thermodynamics of a quantum field in a background with boundaries one usually applies a global procedure which consists first in calculating the Helmholtz free energy $F$ \cite{dow78,amb83,kir91a,kir91b,lim07,gey08,lim09}. Then the internal energy $U$ of the field in a volume $V$ follows from $F$ in the usual manner,  
\begin{equation}
U=\left(\frac{\partial(\beta F)}{\partial \beta}\right)_{V},
\label{fu}
\end{equation}
where $\beta:=1/k_{B}T$.
It should be remarked that in the implementation of this global procedure zeta function is a mathematical tool very much used.

The study of connections between thermodynamics of fields in backgrounds
with boundaries  and  the local ensemble average $\left<T^\mu{}^\nu\right>$ goes back to the 1960's and 1970's \cite{bro69,dow77,bal78,dow78,unw79}. Such a study has been spoiled by the fact that typically
$\left<T^\mu{}^\nu\right>$ has a nonintegrable divergence at idealized boundaries \cite{bal78,deu79},
where idealized boundary conditions are assumed to hold in order to model  real walls of a cavity. Various proposals to ``regularize'' such divergences have appeared in the literature 
\cite{ken80,ken82,for98,rom02,gra04,ful07,ful10,mil11,bou12,maz11,bar12,mil14}
and many aspects of the subject is still actively investigated \cite{mur16}.
The present work wishes to examine  further the connections mentioned above.

The rest of the article is organized as follows. The next section is devoted to reviewing 
the calculation in ref. \cite{mor15} of
$\left<T^\mu{}^\nu\right>$ for a massless scalar field in $N$-dimensional Minkowski spacetime, where space is divided in two halves by a Dirichlet or by a Neumann plane wall
(two models of reflecting wall). Some aspects of the inhomogeneous and anisotropic 
$\left<T^\mu{}^\nu\right>$ are explored further for the purpose of use in the following sections.
In section \ref{energy}, the local procedure to determine the internal energy
is applied. Such a calculation of $U$ is performed when $N>2$ and when $N=2$,
and the results are compared with those in the literature, for related backgrounds,
where global procedures to obtain $U$ have been used.  
In section \ref{freeenergy}, the expressions found for $U$ in the previous section are
used to calculate the free energy $F$ by integrating eq. (\ref{fu}). Once one knows $F$,
familiar thermodynamic relations lead to the entropy and two thermodynamic pressures
(there are two different thermodynamic pressures due to anisotropy). The relationship
between the thermodynamic pressures and the stress components of $\left<T^\mu{}^\nu\right>$ is also established in section \ref{freeenergy}. Section \ref{comments} closes the
article, presenting a summary and conclusions.
(In the rest of the text, $k_{B}=\hbar=c=1$.)

\section{Stress-energy-momentum tensor ensemble average}
\label{stress}
Before addressing $\left<T^\mu{}^\nu\right>$ in the presence of a reflecting plane wall
in flat spacetime it is pertinent to review the tools that allow calculation of 
$\left<T^\mu{}^\nu\right>$ in arbitrary backgrounds. Conventions and the material of the review below are
those in the textbooks \cite{dav82,ful89}, and have been used in ref. \cite{mor15}.
\subsection{Review}
\label{review}
One begins with  an $N$-dimensional
spacetime with metric tensor $g_{\mu\nu}(x)$, where
$\mu$ in the generalized coordinate $x^{\mu}$ runs from 0 to $N-1$ 
and $x^{0}$ is identified with the time coordinate $t$
as is usually done. 
In the Heinsenberg picture
a massless free scalar field is described by the operator
$\phi(t,{\bf x})$ satisfying the covariant
equation
\begin{equation}
\left(g^{\mu\nu}\nabla _{\mu}\nabla _{\nu}+\xi R\right)\phi(x)=0,
\label{se}
\end{equation}
where 
$\xi$ is a dimensionless parameter multiplying
the curvature scalar $R$ and for this reason called curvature coupling parameter. 
It should be noted that when spacetime is locally flat (which is the case of interest in this paper) 
$R=0$, then choosing flat coordinates such that $g^{\mu\nu}=\eta^{\mu\nu}$ and $\nabla _{\mu}=\partial_{\mu}$,
eq. (\ref{se})  becomes the well known wave equation.
When 
$\xi$ equals
\begin{equation}
\xi_{N}:=\frac{N-2}{4(N-1)},
\label{cc}
\end{equation}
it follows that the field eq. (\ref{se}) is 
covariant under conformal transformations, i.e.,
$g_{\mu\nu}\rightarrow\Omega^{2}g_{\mu\nu}$
and $\phi\rightarrow\Omega^{(2-N)/2}\phi$, where 
$\Omega(x)$ is a real function. In the literature the conformal coupling corresponds
to $\xi=\xi_{N}$ and the minimal coupling to $\xi=0$.

To complete the scheme
one must specify the quantum commutation
relations
\begin{eqnarray}
&&\quad\left[\phi(t,{\bf x}),\phi(t,{\bf x}')\right]=0,
\qquad\left[\pi(t,{\bf x}),\pi(t,{\bf x}')\right]=0,
\nonumber\\
&&\quad\qquad\qquad\left[\phi(t,{\bf x}),\pi(t,{\bf x}')\right]=
i\delta\left({\bf x}-{\bf x'}\right),
\label{scr}
\end{eqnarray}
where $\pi=\sqrt{|g|}g^{0\nu}\partial_{\nu}\phi$.
A background with nontrivial topology may be taken
into account by considering boundary conditions.

The Hilbert space is constructed
from eqs. (\ref{se}) and (\ref{scr})
as in Minkowski coordinates; but with the crucial difference
that the normal modes are not plane waves.
Once the modes are determined from eq. 
(\ref{se}) and boundary conditions, a vacuum state $|0\rangle$ is defined
as the state that vanishes under the action of the 
annihilation operator.
The normal modes (and consequently the vacuum)
may or may not be associated with some symmetry of
the background (e.g., in globally flat spacetime, plane waves
are associated with translation invariance).
This makes the particle concept less useful for
dealing with quantum fields
in nontrivial backgrounds since there is no
privileged vacuum, i.e., a particular
symmetry. Therefore 
one looks at expectation values of observables instead. 
A simple example is the
vaccum fluctuation $\langle 0|\phi ^{2}(x)|0\rangle$,
which can be formally written as 
\begin{equation}
\langle\phi ^{2}(x)\rangle=
\lim_{x'\rightarrow x}\frac{1}{2}G^{(1)}(x,x'),
\label{vev1}
\end{equation}
where $G^{(1)}(x,x')/2$ is obtained by symmetrizing 
$\langle 0|\phi(x)\phi(x')|0 \rangle$, i.e.,
\begin{equation}
G^{(1)}(x,x'):=\langle 0|\{\phi(x),\phi(x')\}|0 \rangle,
\label{hadamard}
\end{equation}
known as Hadamard function. Note that eq. (\ref{hadamard})
also satisfies eq. (\ref{se}).
When the limit in eq. (\ref{vev1}) is taken
a divergent quantity arises. If  spacetime is locally
flat (again, which is the case of interest in this paper),
divergences are cured simply by removing the corresponding contribution
in Minkowski spacetime. 
For nonlocally flat spacetimes
more elaborate renomalization procedures must be applied.

The Feynman propagator is defined as usual, i.e.,  as 
the vacuum expectation value of a time-ordered product,
\begin{equation}
iG_{{\cal F}}(x,x'):=
\langle 0|{\cal T}\left(\phi(x)\phi(x')\right)|0\rangle.
\label{sfp}
\end{equation}
This definition implies that
the imaginary part of $G_{{\cal F}}$ is given by
$-G^{(1)}/2$. Thus eq. (\ref{vev1}) may be recast as
\begin{equation}
\langle\phi ^{2}(x)\rangle=
i\lim_{x'\rightarrow x}G_{{\cal F}}(x,x'),
\label{vev2}
\end{equation}
where it has been used the fact that the real part
of the Feynman propagator plays no role in the
final result. By considering  that the
derivative of the step-function (present in the
definition of the time-ordered product) is a 
$\delta$-function and by observing eqs. (\ref{se}) and
(\ref{scr}) it follows that the Feynman
propagator is a Green function of the generalized wave equation,
namely,
\begin{equation}
\left(g^{\mu\nu}\nabla _{\mu}\nabla _{\nu}+\xi R\right)G_{{\cal F}}(x,x')=
-\frac{1}{\sqrt{|g|}}\delta(x-x').
\label{fpe}
\end{equation}
Therefore one solves 
eq. (\ref{fpe}) and then uses eq. (\ref{vev2}) to
find the vacuum fluctuation.

In order to address the energy and momentum content of
the field the following differential operator is defined,
\begin{equation}
{\cal D}_{\mu\nu}(x,x'):=
(1-2\xi)\nabla_{\mu}\nabla_{\nu '}
+(2\xi-1/2)g_{\mu\nu}
\nabla_{\sigma}\nabla^{\sigma '}
-2\xi\nabla_{\mu}\nabla_{\nu },
\label{dop}
\end{equation}
where the prime in $\nabla_{\nu '}$ indicates
that the covariant derivative is 
taken with respect to $x^{\nu'}$.
By letting  eq. (\ref{dop}) act on $G^{(1)}/2$
and considering  the field equation (\ref{se}) it results that
\begin{equation}
\langle T_{\mu\nu}(x)\rangle=
i\lim_{x'\rightarrow x}{\cal D}_{\mu\nu}(x,x')
G_{{\cal F}}(x,x'),
\label{st}
\end{equation}
where
(semicolons below denote covariant differentiation)
\begin{equation}
T_{\mu\nu}(x):=\left(1/2-\xi\right)
\left(\phi_{;\mu}\phi_{;\nu}+\phi_{;\nu}\phi_{;\mu}\right)
+\left(2\xi-1/2\right)g_{\mu\nu}\phi_{;\sigma}\phi^{;\sigma}
-\xi\
\left(\phi_{;\nu\mu}\phi+\phi\phi_{;\nu\mu}\right).
\label{sto}
\end{equation}
An  examination of the operator defined in eq. (\ref{sto}),
where the field and its derivatives appear
conveniently symmetrized, reveals that
it corresponds to the flat spacetime expression for the 
classical stress-energy-momentum tensor,
i.e., $2|g|^{-1/2}\delta S/\delta g^{\mu\nu}$, where
$S=\int{\cal L}dx^{N}$ is the action functional and 
\begin{equation}
{\cal L}=\frac{1}{2}\sqrt{|g|}\left[
g^{\mu\nu}\phi_{;\mu}\phi_{;\nu}
-\xi R\phi^{2}\right]
\label{slag}
\end{equation}
is the Lagrangian density associated with 
eq. (\ref{se}) through the least action principle,
$\delta S=0$.
The expression in eq. (\ref{st}) is by
definition the vacuum expectation value of the 
stress-energy-momentum tensor 
of a massless scalar field in (locally) flat spacetime
[the corresponding expression in curved spacetime may be
obtained by suitably inserting terms depending
on the curvature in eq. (\ref{dop})].

It should be remarked that in flat geometry $\xi$ does 
not appear either in eq. (\ref{slag}) or, consequently,
in eq. (\ref{se}) since $R=0$. However it does appear in the expression
for the stress-energy-momentum tensor [cf. eq. (\ref{sto})] and the reason to do so is that
the variation with respect to the metric tensor,
$\delta S/\delta g^{\mu\nu}$, is made before 
solving Einstein's  equations which yield
the flat geometry.

Another point worth mentioning here is that 
the only dependence on $\xi$ of 
the (classical) energy density 
$T_{00}(x)$ in (locally) flat spacetime appears in a term 
which is a spatial divergence
(colons below denote differentiation with respect to Cartesian coordinates),
\begin{equation}
-2\xi(\phi\phi,_{j}),_{j}.
\label{divergence}
\end{equation}
Thus by integrating $T_{00}(x)$
over space eq. (\ref{divergence}) does not contribute
if $\phi$ (or its derivative) vanishes at the boundaries of the
spatial region, resulting that all values of $\xi$ should lead
to the same total energy.

In other to introduce thermal effects in
quantum field theory one borrows tools and concepts
from quantum statistical mechanics.
A quantum system in thermodynamical 
equilibrium at temperature $1/\beta$ is characterized by the 
Hamiltonian operator $H$ and by the density matrix (operator) $\varrho$. 
The functional form of $\varrho$ is obtained by requiring that the entropy
$S:=-{\rm Tr}\varrho\log\varrho$
is a maximum, leading to 
\begin{equation}
\varrho=\frac{e^{-\beta H}}{Z}.
\label{dmatrix}
\end{equation}
In order to fulfill its probabilistic interpretation
$\varrho$ should satify ${\rm Tr} \varrho=1$.
Consequently the partition function $Z$ in eq. (\ref{dmatrix}) equals
${\rm Tr}\ \exp\{-\beta H\}$ and the thermal average of an observable $A$
is given by 
\begin{equation}
\langle A\rangle_{\beta}={\rm Tr}\varrho A,
\label{taverage}
\end{equation}
which at zero temperature ($\beta=\infty$) reduces to the vacuum average. 

Next 
a scalar field  $\phi(x)$ is taken to be in thermodynamical
equilibrium with a reservoir at temperature $1/\beta$.
For a time independent Hamiltonian the field evolves according to
\begin{equation}
\phi(t,{\bf x})=U(t_{0}-t)\phi(t_{0},{\bf x})U(t-t_{0}),
\label{tevolution}
\end{equation}
where 
$U(t):=\exp\{-iHt\}$ 
is the familiar time evolution operator.
The thermal Feynman propagator is still defined by eq. (\ref{sfp})
with the vacuum average replaced by the thermal average according to eq. 
(\ref{taverage}). Noting the 
cyclic property of the trace, eqs. 
(\ref{dmatrix}),
(\ref{taverage}) and (\ref{tevolution}) lead to 
$G_{\cal F}(t)=G_{\cal F}(t+i\beta)$ (other
coordinates have been omited). As the propagators are
analytic functions of imaginary values of the time $t$,
one can analytically
continue $\tau:=it$ to real values and the boundary condition
above becomes
\begin{equation}
G_{\cal F}^{(\beta)}(\tau)=G_{\cal F}^{(\beta)}(\tau+\beta),
\label{tfeynman}
\end{equation}
where $G_{\cal F}^{(\beta)}(\tau):=G_{\cal F}(-i\tau)$.
Then, once
$G_{\cal F}^{(\beta)}(\tau)$ is determined by solving eq. (\ref{fpe}) with 
eq. (\ref{tfeynman}) satisfied, 
one can analytically continue back to real values of $t$
and interpret the effects at temperature $T=1/\beta$
[see, e.g., eqs. (\ref{vev2}) and (\ref{st})].

\subsection{$\left<T^\mu{}^\nu\right>$ in the presence of a reflecting plane wall}
\label{wall}
Consider a Minkowski spacetime with $N\geq 2$ dimensions where points are labelled by usual flat coordinates,
$(t,x,y,z,\cdots)$. An arbitrarily large plane wall is taken at $x=0$ simulating a real wall of a large cavity in which a massless neutral scalar field $\phi$ is in thermodynamic equilibrium at temperature $T$. 
In order to model a reflecting wall at $x=0$, 
the thermal Feynman propagator 
is assumed to satisfy the Dirichlet or the Neumann boundary conditions
besides satisfying eq. (\ref{tfeynman}) 
(for more details regarding this material see ref. \cite{mor15}). 
Only the diagonal components of the corresponding ensemble average of the 
stress-energy-momentum tensor are nonvanishing \cite{mor15}, i.e.
[it should be noted that $\left<T^\mu{}^\nu\right>$ follows from eqs. (\ref{dop}) and (\ref{st}) by raising indices],
\begin{equation}
\left<T^\mu{}^\nu\right>={\rm diag}(\rho,P_{\perp},P_{\parallel},\cdots,P_{\parallel}).
\label{emt}
\end{equation} 

The energy density in eq. (\ref{emt}) is given by,
\begin{equation}
\rho=
\rho_{{\tt vacuum}}
+
\rho_{{\tt mixed}}
+
\rho_{{\tt thermal}},
\label{energy-density}
\end{equation} 
where
\begin{eqnarray}
&&\rho_{{\tt vacuum}}=\pm\frac{2(N-1)}{(2\pi^{1/2})^N}(\xi-\xi_{N})\Gamma\left(\frac{N}{2}\right)x^{-N},
\nonumber
\\
&&\rho_{{\tt mixed}}
=\pm\frac{4}{\pi^{N/2}}\Gamma\left(\frac{N}{2}\right)T^{N}
\sum_{m=1}^{\infty}\left[(2Tx)^2+m^2\right]^{-(N+2)/2}
\nonumber
\\
&&\hspace{2.0cm}\times\left[(N-1)(\xi-\xi_N)\left(4T^2x^2+m^2\right)-\xi Nm^2\right],
\nonumber
\\
&&\rho_{{\tt thermal}}=\frac{N-1}{\pi^{N/2}}\Gamma\left(\frac{N}{2}\right)\zeta(N)T^{N},
\label{densities}
\end{eqnarray}
upper signs for Dirichlet and lower signs for Neumann (as in the rest of the text).
When $\xi=\xi_{N}$ [cf. eq. (\ref{cc})]
one has that $\left<T^\mu{}^\nu\right>$ is traceless. 
It is worth noting that as $T\rightarrow 0$ only the first term in eq. (\ref{energy-density}) is left behind. As $x\rightarrow\infty$ only the Planckian 
$\rho_{{\tt thermal}}$ remains. (In all formulas $x$ is taken to be nonnegative which corresponds to choosing one side of the wall.)
The contribution $\rho_{{\tt mixed}}$ in eq. (\ref{energy-density})
bridges the vacuum $\rho_{{\tt vacuum}}$ and the blackbody 
$\rho_{{\tt thermal}}$ contributions and that is, in fact, a general feature common to all densities \cite{mor15}.


The remaining components of 
$\left<T^\mu{}^\nu\right>$ in eq. (\ref{emt}) are,
\begin{equation}
P_{\perp}=\frac{\rho_{{\tt thermal}}}{N-1},
\hspace{1.0cm}
P_{\parallel}=
-\rho_{{\tt vacuum}}+P_{\parallel\ {\tt mixed}}+
\frac{\rho_{{\tt thermal}}}{N-1},
\label{stresses}
\end{equation}
where
\begin{samepage}
\begin{eqnarray}
P_{\parallel\ {\tt mixed}}
=-\frac{4}{\pi^{N/2}}\Gamma\left(\frac{N}{2}\right)T^{N}
\sum_{m=1}^{\infty}\left[(2Tx)^2+m^2\right]^{-(N+2)/2}\hspace{2cm}&&
\nonumber
\\
\times\left[(N-1)(\xi-\xi_N)\left(4T^2x^2+m^2\right)-(\xi-1/4) Nm^2\right].&&
\label{mixed11}
\end{eqnarray}
\end{samepage}
It follows from eqs. (\ref{stresses}) and (\ref{mixed11})
that $P_{\perp}\neq P_{\parallel}$ and that isotropy holds only
asymptotically, i.e., when $x \rightarrow\infty$.
The following relation between 
$\rho_{{\tt mixed}}$ and $P_{\parallel\ {\tt mixed}}$
will be useful later in the paper,
\begin{eqnarray}
\rho_{{\tt mixed}}-(N-2)P_{\parallel\ {\tt mixed}}
=\frac{4}{\pi^{N/2}}(N-1)(\xi-\xi_N)\Gamma\left(\frac{N}{2}\right)T^{N}&&
\nonumber
\\
\hspace{2.0cm}
\times\sum_{m=1}^{\infty}
\left[(N-1)(2Tx)^2-m^2\right]\left[(2Tx)^2+m^2\right]^{-(N+2)/2}.
&&
\label{relation}
\end{eqnarray}

As has been shown in ref. \cite{del15} there is a range over which the values of $\xi$ are restricted
in order that stable thermodynamic equilibrium prevails. For the Dirichlet boundary condition
$\xi$ must satisfy,
\begin{eqnarray}
\frac{N-3}{4(N-2)}<\xi<\frac{1}{4}, 
\hspace{1.5cm} 
&&N>2,
\label{dboundary1}
\\
\xi<\frac{1}{4}, 
\hspace{1.5cm} 
&&N=2,
\label{dboundary2}
\end{eqnarray}
and for the Neumann boundary condition $\xi$ must be such that,
\begin{eqnarray}
\frac{3-2N}{4}<\xi<\frac{3N-5}{4(N-2)}, 
\hspace{1.5cm} 
&&N>2,
\label{nboundary1}
\\
\xi>-\frac{1}{4}, 
\hspace{1.5cm} 
&&N=2.
\label{nboundary2}
\end{eqnarray}
It should be pointed out that the conformal coupling in eq. (\ref{cc}) fits
these bounds; but that the minimal coupling ($\xi=0$) does not fit in eq. (\ref{dboundary1}) for $N>3$. Although equalities have not been included 
in the bounds of $\xi$
in eqs. (\ref{dboundary1}) to (\ref{nboundary2}) there are evidences that they
still hold when equalities are included. Before closing this section it should also be remarked that eqs. (\ref{nboundary1}) and (\ref{nboundary2}) were not presented in ref. \cite{del15} for arbitrary $N$, but the way to obtain them is the same as that leading to eqs. (\ref{dboundary1}) and (\ref{dboundary2}) which appeared first
in ref. \cite{del15}.

\section{Internal energy}
\label{energy}

When $N>2$, one assumes that the reflecting plane wall of the large cavity described in the previous section has area $A$. The cavity is considered to be cylindrical with height $a$, such that its volume is $Aa$. Only boundary conditions on this particular 
wall will be taken into account, although connection with other configurations of
walls in the literature will be made.
The internal energy $U$ of the scalar radiation in the cavity is obtained
by integrating eq. (\ref{energy-density}) and noting eq. (\ref{densities}),
\begin{equation}
U=U_{{\tt vacuum}}+U_{{\tt mixed}}+U_{{\tt thermal}},
\label{cavity-energy}
\end{equation}
where,
\begin{eqnarray}
&&U_{{\tt vacuum}}=A\int_{\epsilon}^{a}\rho_{{\tt vacuum}}\, dx,\hspace{1.0cm}U_{{\tt mixed}}=A\int_{0}^{a}\rho_{{\tt mixed}}\, dx,
\label{vacuum-mixed}
\end{eqnarray}
and,
\begin{equation}
U_{{\tt thermal}}=\frac{N-1}{\pi^{N/2}}\, Aa\,\Gamma\left(\frac{N}{2}\right)\zeta(N)T^{N}.
\label{thermal}
\end{equation}
The nonvanishing lower bound of the integration in eq. (\ref{vacuum-mixed}), 
$\epsilon$, takes into account the nonintegrable divergence mentioned in section \ref{int}. When $N=2$, one simply sets $A$ equal to unity in the arguments above, including eqs. (\ref{cavity-energy}) to (\ref{thermal}). It should be
noticed that eq. (\ref{thermal}) is the usual internal energy of a gas of massless scalar bosons in an $(N-1)$-dimensional cavity of volume $Aa$. In particular, for $N=4$,
eq. (\ref{ie})	is reproduced.

In the present calculation the cavity is considered arbitrarily large in the sense
that for any nonvanishing $T$, it is assumed that 
\begin{equation}
Ta\gg 1.
\label{htemperature}
\end{equation}
Consequently, one can formally take $a=\infty$ in integrations in eq. (\ref{vacuum-mixed}) and,  
in doing so, the result is as if only leading contributions in each integration were kept. In dealing with the limit $\epsilon\rightarrow 0$ in 
eq. (\ref{vacuum-mixed}) one of the regularization schemes  mentioned previously must be used. It turns out that any of them yields \cite{ken80,amb83,for98,rom02,ful07},
\begin{equation}
\int_{0}^{\infty}\rho_{{\tt vacuum}}\, dx=0,
\label{renormalization}
\end{equation} 
and thus eq. (\ref{cavity-energy}) becomes,
\begin{equation}
U=U_{{\tt thermal}}+U_{{\tt mixed}},
\label{renormalized-energy}
\end{equation}
with
\begin{equation}
U_{{\tt mixed}}=A\int_{0}^{\infty}\rho_{{\tt mixed}}\, dx.
\label{mixed}
\end{equation}
Note that $U_{{\tt mixed}}$ in eq. (\ref{renormalized-energy}) should be seen as the leading correction to the Planckian $U_{{\tt thermal}}$ in eq. (\ref{thermal}) due to the presence of a reflecting wall. Terms in eq.  (\ref{renormalized-energy}) have been rearranged to denote this fact. In order to proceed with the calculation of $U$, the cases $N>2$ and $N=2$
will be treated separately.
\subsection{\emph{N}$>$2} 
\label{n>2}

Before using in eq. (\ref{mixed}) the expression for $\rho_{{\tt mixed}}$ 
as given in eq. (\ref{densities}), it is convenient to consider the definitions,
\begin{eqnarray}
&&
f(\alpha):=\frac{1}{2}\int_{0}^{\infty}d\chi\sum_{m=1}^{\infty}
\left[(\alpha-1)\chi^2-m^2\right]\left(\chi^2+m^2\right)^{-(\alpha+2)/2},
\label{f}
\end{eqnarray}
and
\begin{eqnarray}
&&
g(\alpha):=\frac{1}{2}\int_{0}^{\infty}d\chi\sum_{m=1}^{\infty}
\left(\chi^2+m^2\right)^{-\alpha}.
\label{g}
\end{eqnarray}
Then, for $\chi:=2Tx$, it follows that
\begin{equation}
U_{{\tt mixed}}=\pm\frac{4A}{\pi^{N/2}}
\Gamma\left(\frac{N}{2}\right)T^{N-1}
\left[\xi f(N)-(N-1)\xi_{N}\, g(N/2)\right], \hspace{1.5cm} N>2.
\label{mixed-energy}
\end{equation}
When $\alpha>2$, by using $t:=\chi^{2}/m^{2}$ in eq. (\ref{f}), it results
\begin{equation}
f(\alpha)=\frac{1}{4}\zeta(\alpha-1)
\int_{0}^{\infty} dt\, t^{-1/2}
\left[(\alpha-1)t-1\right]\left(1+t\right)^{-(\alpha+2)/2}.
\label{mf}
\end{equation}
Noticing now that,
$$\int dt\, t^{-1/2}
\left[(\alpha-1)t-1\right]\left(1+t\right)^{-(\alpha+2)/2}=-2t^{1/2}(1+t)^{-\alpha/2},$$
eq. (\ref{mf}) yields
\begin{equation}
f(\alpha)=0, \hspace{1.5cm} \alpha>2. 
\label{nulef}
\end{equation}
Thus, eq. (\ref{mixed-energy}) agrees with the common knowledge that total energies in flat spacetime do not depend on $\xi$ [see, e.g., ref. \cite{ful03} and the paragraph containing eq. (\ref{divergence})].
Function $g(\alpha)$ in eq. (\ref{g}) can be manipulated in a similar manner as $f(\alpha)$, following that (using, e.g., \textsc{Mathematica} {\cite{mat17}),
\begin{equation}
g(\alpha)=\frac{\pi^{1/2}}{4}\zeta(2\alpha-1)\frac{\Gamma(\alpha-1/2)}{\Gamma(\alpha)}, \hspace{1.5cm} \alpha>1.
\label{finalg}
\end{equation}
By inserting eqs. (\ref{nulef}) and (\ref{finalg}) in eq. (\ref{mixed-energy}), one ends up with
\begin{equation}
U_{{\tt mixed}}=\mp\frac{1}{4}
\frac{N-2}{\pi^{(N-1)/2}}\, A\,\Gamma\left(\frac{N-1}{2}\right)\zeta(N-1)T^{N-1},
\label{final-mixed-energy}
\end{equation}
where eq. (\ref{cc}) has been used. It should be stressed that eq. (\ref{final-mixed-energy}) holds when $N>2$.

Comparing eqs. (\ref{thermal}) and (\ref{final-mixed-energy}),
it is seen that the contribution in eq. (\ref{renormalized-energy})
due to the presence of the plane Dirichlet wall is a deficit of $1/4$
of  blackbody energy in a $(N-2)$-dimensional cavity;
and a excess of $1/4$ for the plane Neumann wall.
According to this loose picture the corresponding ``bosons'' can be thought to be confined to the wall.

Working with zeta function regularization in static spacetimes,
ref. \cite{dow84} calculated the high temperature behaviour of the free energy of a massless scalar field confined to an $(N-1)$-dimensional compact space ${\cal M}$,
whose volume is $|{\cal M}|$, and where the area of the corresponding boundary $\partial{\cal M}$ is $|\partial{\cal M}|$, finding,
\begin{eqnarray}
F=
-\frac{1}{\pi^{N/2}}\left[|{\cal M}|\Gamma\left(\frac{N}{2}\right)\zeta(N)T^{N}\mp\frac{1}{4}\pi^{1/2}|\partial{\cal M}|\Gamma\left(\frac{N-1}{2}\right)\zeta(N-1)T^{N-1}
\right],&&
\label{dowker}
\end{eqnarray}
where subleading contributions involving spacetime curvature have been omitted. Now, using eq. (\ref{dowker}) in eq. (\ref{fu}), it results  eq. (\ref{renormalized-energy}) with $|{\cal M}|$ in the place of $Aa$ in eq. (\ref{thermal}), and $|\partial{\cal M}|$ 
in the place of $A$ in eq. (\ref{final-mixed-energy}).
Such a correspondence suggests that, at high temperature $T$, each reflecting wall of a large cavity
contributes with  $U_{{\tt mixed}}$ in eq. (\ref{final-mixed-energy}). Indeed, for a slab, which consists of two parallel plane walls separated by a distance $a$,
in the regime where eq. (\ref{htemperature}) holds it follows that the leading correction to $U_{{\tt thermal}}$ in eq. (\ref{thermal})
is twice eq. (\ref{final-mixed-energy}) \cite{amb83,lim07}.

\subsection{{\emph N}=2}
\label{n=2}

When $N=2$, noticing  eq. (\ref{thermal}), one goes back to eq. (\ref{renormalized-energy}) and writes,
\begin{equation}
U=\frac{\pi}{6}aT^{2}+U_{{\tt mixed}},
\label{n=2renormalized-energy}
\end{equation}
where $U_{{\tt mixed}}$ is given by eq. (\ref{mixed}), now with
$A\equiv 1$  and $\rho_{{\tt mixed}}$ given in eq. (\ref{densities})
for $N=2$,
\begin{equation}
U_{{\tt mixed}}=\pm\frac{4}{\pi}T\xi f(2).
\label{2mixed}
\end{equation}
Noting eq. (\ref{f}) and using ref. \cite{mat17}, it results,
\begin{eqnarray}
f(2)&=&\frac{1}{2}\int_{0}^{\infty}d\chi\sum_{m=1}^{\infty}
\frac{\chi^2-m^2}{(\chi^2+m^2)^{2}}
\nonumber
\\
&=&\frac{1}{2}\int_{0}^{\infty}d\chi\frac{(\pi\chi)^{2}{\rm csch}^{2}(\pi\chi)-1}{2\chi^{2}}=-\frac{\pi}{4},
\label{f2}
\end{eqnarray}
which should be compared with eq. (\ref{nulef}). 
[See also in figure \ref{figure} plots against $\chi$ of the integrands of $f(3)$ in eq. (\ref{f}) and of $f(2)$ in eq. (\ref{f2}).]           
Using eq. (\ref{f2}) in eq. (\ref{2mixed}), one finds, unexpectedly,  that the correction to the blackbody contribution in eq. (\ref{n=2renormalized-energy}) depends on $\xi$, namely,
\begin{equation}
U_{{\tt mixed}}=\mp\xi T.
\label{planckian-correction}
\end{equation}

\begin{figure}[h]
\center
\includegraphics[scale=0.7]{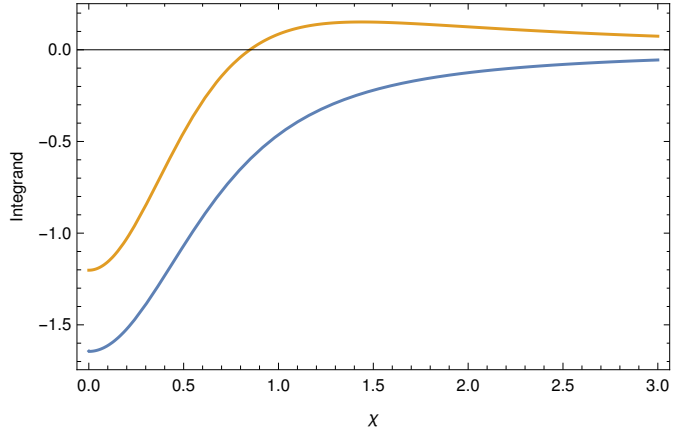}
\caption{\footnotesize{The upper plot (yellow) is the the integrand of $f(3)$ and the lower plot (blue) is the integrand of $f(2)$. The upper plot initially runs under zero, and then over zero, explaining eq. (\ref{nulef}). The lower plot always
runs under zero, such that eq. (\ref{f2}) holds.}}
\label{figure}
\end{figure}

Looking at the literature for related configurations of reflecting ``walls'' (more properly, reflecting points) in spacetimes with $N=2$ dimensions, one sees that the general formula in eq. (\ref{dowker}) is not applicable. Nevertheless, refs. \cite{amb83} and \cite{lim09}, by using global regularization procedures appropriate for $N=2$, have arrived to an internal energy given by, $\pi a T^{2}/6 -T/2$,
for the case of a slab of width $a$ 
regardless the boundary condition
is Dirichlet's or Neumann's. Now, taking into account eqs. (\ref{n=2renormalized-energy}) and (\ref{planckian-correction}),
and assuming that (as when $N>2$) the correction to the
Planckian internal energy in the slab is twice eq. (\ref{planckian-correction}), 
the result of the  
global procedure is reproduced if $\xi$ is taken to be $1/4$ for the Dirichlet, and $-1/4$ for the Neumann boundary conditions. It should be remarked that these values fit the bounds on $\xi$ given in the previous section [note that, as mentioned earlier, equalities are likely not discarded in eqs. (\ref{dboundary2}) and (\ref{nboundary2})].

At this point a word of caution should be given regarding the assumption above that the correction to the
Planckian internal energy in the slab is twice eq. (\ref{planckian-correction}). In the case of a slab (two reflecting walls separated by $a$), besides  $\rho_{{\tt mixed}}$
given in eq. (\ref{densities}) for $N=2$, additional terms such as  $\pm \xi T/a$ might appear. Consequently, recalling eq. (\ref{planckian-correction}), the integration in eq. (\ref{vacuum-mixed}) would sweep away the dependence on $\xi$ in the expression of $U_{{\tt mixed}}$.
Although this is certainly a possibility which would lead to agreement with global procedure \cite{amb83,lim09}, only a calculation of
$U_{{\tt mixed}}$ using the present local procedure extended to 
the case of two reflecting walls could give an answer for sure
\footnote{This author intends to investigate this point and results will appear elsewhere.}.
 
\section{Other thermodynamic quantities}
\label{freeenergy}

Before studying other thermodynamic quantities in the large cavity, 
it is convenient at this stage to consider the space averages
(integration over space divided by the volume),
$p_{\perp}$ and $p_{\parallel}$, of
$P_{\perp}$ and $P_{\parallel}$ 
in eq. (\ref{emt}),
\begin{equation}
p_{\perp}=\frac{1}{\pi^{N/2}}\Gamma\left(\frac{N}{2}\right)\zeta(N)T^{N},
\hspace{1.0cm}
p_{\parallel}=
p_{\perp}+
\frac{1}{a}\int_{0}^{\infty}P_{\parallel\ {\tt mixed}}\,  dx,
\label{pressures}
\end{equation}
where eqs. (\ref{densities}), (\ref{stresses}) and (\ref{renormalization}) have been used.
Note that, naturally, the formula for $p_{\parallel}$ only applies when $N>2$.
As will be seen below, thermodynamics involves these space averages.


\subsection{{\emph N}$>$2} 
\label{2n>2}

By integrating eq. (\ref{relation}) over $x$ noting eq. (\ref{f}), one finds that its right hand side yields a term proportional to $f(N)$
which, on account of eq. (\ref{nulef}), vanishes. Using then eq. (\ref{final-mixed-energy}), it follows that $p_{\parallel}$ in eq. (\ref{pressures}) is given by,
\begin{equation}
p_{\parallel}=
p_{\perp}\mp
\frac{1}{4a\pi^{(N-1)/2}}\, \Gamma\left(\frac{N-1}{2}\right)\zeta(N-1)T^{N-1}.
\label{ppressure}
\end{equation}

Considering $T$, $A$ and $a$ as independent thermodynamic parameters
and observing eqs. (\ref{renormalized-energy}), (\ref{thermal}) and (\ref{final-mixed-energy}), integration of eq. (\ref{fu}) leads
to the free energy in eq. (\ref{dowker}) with 
$|{\cal M}|$ and $|\partial{\cal M}|$ replaced by 
$Aa$ and $A$, respectively [cf. text after eq. (\ref{dowker})].
It can be readily checked that $F$ so obtained satisfies, 
\begin{equation}
p_{\perp}=-\frac{1}{A}\left(\frac{\partial F}{\partial a}\right)_{T,A},
\hspace{1.0cm}
p_{\parallel}=-\frac{1}{a}\left(\frac{\partial F}{\partial A}\right)_{T,a},
\label{thermodynamic-pressures}
\end{equation}
i.e., the thermodynamic pressures are the space averages in eqs.
(\ref{pressures}) and (\ref{ppressure}). 
Note that the pressure perpendicular to the wall $p_{\perp}$ is the familiar blackbody pressure [by setting  $N=4$ in eq. (\ref{pressures}),  $p_{\perp}$ yields $p$ in eq. (\ref{gas})],
and that the pressure parallel to the wall $p_{\parallel}$ is diminished or increased  by the presence of a Dirichlet or Neumann wall, respectively [see text in the paragraph starting after eq. (\ref{final-mixed-energy})]. At this point, it should be added that
an integration constant (independent of $T$) in $F/T$ has been ignored
when integrating eq. (\ref{fu}). If such a constant is taken to be different from zero, the partial derivatives in eq. (\ref{thermodynamic-pressures}) could give rise to Casimir terms (linear on $T$ \cite{amb83}) in conflict
with eqs. (\ref{pressures}) and (\ref{ppressure}) where Casimir pressures
are not present (recall that in this paper boundary conditions are  considered on a single wall).

Clearly one can go further to obtain other thermodynamic quantities, for example entropy, 
\begin{equation}
S=-\left(\frac{\partial F}{\partial T}\right)_{A,a},
\label{entropy}
\end{equation}
whose expression follows immediately from eq. (\ref{dowker}).

\subsection{{\emph N}=2}
\label{2n=2}

Turning now to $N=2$, using eqs. (\ref{n=2renormalized-energy}) and (\ref{planckian-correction}) in eq. (\ref{fu}), integration yields,
\begin{equation}
F=-\frac{\pi}{6}aT^{2}\pm\xi T\ln\, (T\ell),
\label{2freeu}
\end{equation}
where $\ell$ is an arbitrary length scale which must be introduced such that the argument of the logarithm is dimensionless. (As done above, an integration constant in $F/T$ has been omitted.) Assuming that $\ell$
is independent of $a$, it follows from eq. (\ref{2freeu}) a thermodynamic pressure which does not depend on $\xi$,
\begin{equation}
p=-\left(\frac{\partial F}{\partial a}\right)_{T}=\frac{\pi}{6}T^{2},
\label{2pressure}
\end{equation}
agreeing with the space average $p_{\perp}$ in eq. (\ref{pressures}). By inserting eq. (\ref{2freeu}) into eq. (\ref{entropy}), it results that,
\begin{equation}
S=\frac{\pi}{3}aT\mp\xi[\ln\, (T\ell)+1],
\label{2entropy}
\end{equation}
i.e., the blackbody entropy is corrected by a term that depends on
$\xi$ and $\ell$. It is worth mentioning that one cannot set $T\rightarrow 0$ in eq. ({\ref{2entropy}}) and argue that entropy diverges as the absolute zero of temperature is approached, since the leading contribution in eq. ({\ref{2entropy}}) is the Planckian form [cf. eq. ({\ref{htemperature})].

Noting eqs. (\ref{2freeu}) and (\ref{2pressure}), it follows that
the Gibbs free energy $G=F+pa$ is given by
\begin{equation}
G=\pm\xi T\ln\, (T\ell),
\label{2gibbs}
\end{equation}
and then only for the minimal coupling $\xi=0$ [which is the same as the conformal coupling when $N=2$, cf. eq. (\ref{cc})] the Euler relation is satisfied. In fact, for $\xi=0$, all thermodynamic quantities reduce to Planckian forms.

\section{Conclusion}
\label{comments}

This paper examined how the single Dirichlet and Neumann plane walls
modify the thermodynamics of standard scalar blackbody radiation.
This was done using the local procedure, which consists in integrating the renormalized energy density to obtain
the internal energy, instead of obtaining the latter by means of the free energy as is usually done following the global procedure.
The main result is that in $N=2$-dimensional spacetime, total thermodynamic quantities depend on the curvature coupling parameter $\xi$ although spacetime is
flat. When  $N>2$, agreement between the local and global procedures was achieved; not so much when $N=2$
due to the presence of $\xi$ in the corresponding total thermodynamic quantities.
The central point of the argument is that the following integration in ``half space''
[see eqs. (\ref{mixed}), (\ref{final-mixed-energy}) and (\ref{planckian-correction})],
$$\int_{0}^{\infty}\rho_{{\tt mixed}}\, dx,$$
does not depend on $\xi$ when $N>2$; but it does depend on $\xi$ when $N=2$.

It should be recalled that although total thermodynamic quantities in $N=2$
carry dependence on the curvature coupling parameter $\xi$, the value of the latter is not completely arbitrary. Namely, eqs. (\ref{dboundary2}) and (\ref{nboundary2}) 
dictate that
$\xi$ must not be greater than $1/4$ in the case of 
Dirichlet's and not smaller than $-1/4$ in the case of Neumann's boundary condition, otherwise stable thermodynamic equilibrium will not prevail.

In dealing with the Dirichlet and the Neumann boundary conditions in the context of flat spacetimes, appearance of $\xi$ in total quantities such as the internal energy [cf. eqs. (\ref{n=2renormalized-energy}) and (\ref{planckian-correction})], the Helmholtz free energy [cf. eq. (\ref{2freeu})], and the entropy [cf. eq. (\ref{2entropy})] is puzzling 
[see paragraph containing eq. (\ref{divergence})]
and asks for further investigation of the thermodynamics of the scalar radiation in the presence of boundaries in two-dimensional backgrounds. As mentioned earlier in the paper this study will appear elsewhere.

Perhaps, it is worth commenting here that there is in literature another instance
where the curvature coupling parameter $\xi$ plays a role in total quantities, even in flat
backgrounds. Namely, in the calculation of geometric entropies \cite{lar96}.


\vspace{1cm}
\noindent{\bf Acknowledgements} -- Work partially supported by
``Funda\c{c}\~{a}o de Amparo \`{a} Pesquisa do Estado de Minas Gerais'' (FAPEMIG).




\end{document}